\documentstyle{article}
\def\lds{|\mkern-2.5mu|}
\def\rds{|\mkern-2.5mu|}

\def\ldc{(\mkern-4mu(}
\def\rdc{)\mkern-4mu)}
\def\ankh{{\dagger\mkern-2mu\dagger}}

\def\Av#1{\left\langle #1\vphantom{\int}\right\rangle}
\def\TO#1{\top\left\lbrack\vphantom{\int}#1\right\rbrack}

\def\mathrm#1{\mbox{#1}}
\def\e{\mbox{e}}
\font\twlmib = cmmib10 scaled \magstep1
\font\tenmib = cmmib10 scaled \magstep0
\def\vec#1{\mbox{\twlmib #1}}
\def\vecs#1{\mbox{\tenmib #1}}
\begin{document}
\title{Thermo Field Dynamics and\\
       Kinetic Coefficients of a \\
       Charged Boson Gas\thanks{Work supported by GSI}}
\author{P.A.Henning \thanks{E-mail address: phenning@tpri6a.gsi.de}
\\
Institut f\"ur Kernphysik der TH Darmstadt and GSI\\[1mm]
        P.O.Box 110552, D-64220 Darmstadt, Germany}
\date{\today}
\maketitle
\begin{abstract}
Thermo Field Dynamics for inhomogeneous systems is generalized
to quantum fields with a continuous single-particle mass spectrum.
The modification of the hamiltonian in states with a local
thermal Bogoliubov symmetry is used to calculate thermal
conductivity and diffusion coefficient of pions interacting
with hot, compressed nuclear matter.
\end{abstract}
\clearpage
\section{Introduction}
Many efforts in contemporary physics are devoted to the study of
relativistic quantum systems in non-equilibrium states, like e.g.
the expansion of hot, compressed nuclear matter. Unfortunately,
the combination of relativistic quantum field theory and
statistical mechanics required for the theoretical description
of such systems, i.e., relativistic transport theory,
is far from well understood and still
subject to discussion. As a result, most publications
in this field either start from field theory and end up with
some transport equation that can hardly be solved -- or solve
a transport equation, which has lost much of its touch with quantum
field theory.

The present paper attempts to bridge this gap through a different
approach: The response of a quantum system to gradients in temperature
and chemical potential is calculated {\em directly\/}, i.e., without
the detour of transport theory. To this end a formalism is introduced,
which incorporates the space-time dependence of
temperature and chemical potential into the quantization rules
for the interacting quantum fields.

The paper is organized as follows: A brief introduction to the
formalism of Thermo Field Dynamics is followed by its application
to a free charged scalar field in global as well as local equilibrium
states. The ideas are then extended to the interacting field,
and used to calculate the full two-point function for
spatially inhomogeneous states approximately.

In section 5, the formal results are then applied to
calculate the currents of the system, which arise due to
the spatial inhomogeneity. Kinetic (or transport) coefficients
are obtained in terms of the spectral function of the interacting
field, and compared to formal results obtained with different methods.
Numerical results are presented for a simple approximation
to pions interacting with a nuclear medium.

An important aspect of finite temperature field theory is
the existence of two different, mutually
commuting representations of the canonical commutation relations:
The Hilbert space of the quantum description is
''doubled'' with respect to the zero temperature case \cite{L88}. For a
description in terms of Green's functions, this requires to use
$2\times2$ matrix valued propagators. The two best developed
realizations of such a formalism are Thermo Field Dynamics
(TFD)\cite{Ubook} and the Closed-Time-Path method \cite{DP91}.

While the vacuum state of quantum field theory is invariant under
Lorentz transformations, matter states have, in general, less
symmetry. This leads to a modification of the particle spectrum
needed for the description of such a state \cite{BS75}.  Hence, a
consistent quantum description requires elementary excitations
with a continuous mass spectrum, rather than physical
quasi-particles of infinite lifetime \cite{L88}.

Formally this is expressed as spectral function of ``particles''
which deviates from a $\delta$-function, i.e., which is more than
a mass-shell constraint.  In several papers it was shown
recently, that in the framework of Thermo Field Dynamics a
physical interpretation of the $2\times 2$ matrix structure is
obtained also for systems with such a nontrivial single-particle
spectrum \cite{YUNA92,hu92,hu92a}. In this sense, TFD is superior
to the Closed Time-Path method.

In ''ordinary'' quantum mechanics,
a thermal state of a quantum system is described by a statistical
operator (or density matrix) $W$, and the measurement of
an observable will yield the average
\begin{equation}\label{av1}
\Av{ {\cal E}(t,\vec{x}) } =
   \frac{\mbox{Tr}\left[ {\cal E}(t,\vec{x})\;W \right]}{
         \mbox{Tr}\left[ W \right]}
\;,\end{equation}
where the trace is taken over the Hilbert space of the quantum system
and ${\cal E}$ is the hermitean operator associated with the
observable. In Thermo Field Dynamics (TFD), the calculation
of this trace is simplified to the calculation of a matrix element
\begin{equation}\label{av3}
\Av{ {\cal E}(t,\vec{x}) } =
\frac{ \ldc W^L \rds \;{\cal E}(t,\vec{x}) \;\lds W^R\rdc}{
       \ldc W^L \rds  W^R\rdc}
\;,\end{equation}
with ''left'' and ''right'' statistical state defined in terms of
the two different commuting representations (see refs.
\cite{Ubook,hu92} for details).

If gradients of the temperature etc. are present in the state,
the hamiltonian of the quantum system, which generates
its time evolution,
contains a part which is due to the spatial inhomogeneity.
The above statistical average in the course of time evolution
will be modified by the gradients present in the system.

If the space-dependence of the temperature and chemical potential
is sufficiently weak, the response of an observable ${\cal E}$
to a gradient is a linear function of the perturbation.
We thus express this
response $\delta\Av{{\cal E}(t,\vec{x})}$ up to first order
in $\widehat{H}_1$ (the gradient part of the hamiltonian) as
\begin{equation}\label{td}
\delta\Av{{\cal E}(t,\vec{x})} = \mathrm{i}\int\limits_{t_0}^t\!\!d\tau
  \Av{\left[{\cal E}(t-\tau,\vec{x}),\widehat{H}_1\right]}
\;.\end{equation}
If the commutator can be calculated, one
can use it to obtain the transport coefficients
of the system, i.e. the coefficients relating the response
to the perturbation. The crucial point however is the commutator:
Its expectation value is generally calculated as
\begin{equation}\label{kubo}
\Av{\left[{\cal E}(t-\tau,\vec{x}),\widehat{H}_1\right]}
= \int\limits_0^\beta\!\!d\lambda \Av{ \frac{\partial}{\partial t}
          {\cal E}(t-\tau-\mathrm{i}\lambda,\vec{x})\,\widehat{H}_1}
\;,\end{equation}
i.e., as the time derivative of a correlation function
with {\em complex time arguments \/}  \cite{HST84,KKER86}.
The above expression is usually referred to as the Kubo formula
\cite{K57}.

The analytical structure of a correlation function  in the
complex time plane is related to the properties of its Fourier
transform in the complex energy plane. However, the latter
is plagued by pathologies of perturbation theory which are
especially severe when considering nuclear phenomena
\cite{TBAB91,h92fock}.
The most natural way to overcome this problem is
to use TFD in a {\em direct\/} computation of the commutator
expectation value of eqn. (\ref{td}).

We therefore consider a complex, scalar boson field
describing spinless charged excitations in a statistical
system not too far from equilibrium. In the spirit of the
first remark, one could think of this field as describing
positive and negative pions in nuclear matter.

According to the reasoning above,
the thermal boson field is then described by two field
operators $\phi_x$, $\widetilde{\phi}_x$  and their adjoints
$\phi_x^\star$, $\widetilde{\phi}_x^\star$, with canonical
commutation relations
\begin{eqnarray}
\left[\phi(t,\vec{x}) , \partial_t
\phi^\star(t,\vec{x}^\prime)\right] & = &
        \mathrm{i}\delta^3(\vec{x}-\vec{x}^\prime) \nonumber \\
\left[\widetilde{\phi}(t,\vec{x}) ,
\partial_t\widetilde{\phi}^\star(t,\vec{x}^\prime)\right] & =-&
         \mathrm{i}\delta^3(\vec{x}-\vec{x}^\prime)
\;\end{eqnarray}
but commuting with each other.
They are combined in a statistical doublet
\begin{equation}\label{dob}
\Phi_x =
\left({\array{c} \phi_x \\\widetilde{\phi}_x^\star
       \endarray}\right)
\;,\end{equation}
and the 2$\times$2 matrix valued propagator of this doublet is
\begin{eqnarray}\label{pd}
D^{(ab)}(x,x^\prime)&  = &
   -\mathrm{i}\ldc W^L\rds \TO{\Phi_x\,\Phi_{x^\prime}^\dagger}
        \lds W^R \rdc/\ldc W^L \rds  W^R\rdc \nonumber \\
   &=&-\mathrm{i}\left( {
   \array{ll}
    \ldc W^L\rds \TO{\phi_x \phi_{x^\prime}^\star} \lds W^R \rdc &
    \ldc W^L\rds \TO{\phi_x\widetilde{\phi}_{x^\prime}} \lds W^R \rdc \\[2mm]
    \ldc W^L\rds \TO{\widetilde{\phi}_x^\star\phi_{x^\prime}^\star} \lds W^R
\rdc &
    \ldc W^L\rds \TO{\widetilde{\phi}_x^\star\widetilde{\phi}_{x^\prime}} \lds
W^R \rdc
   \endarray} \right)\frac{1}{\ldc W^L \rds  W^R\rdc}
\;.\end{eqnarray}
The free as well as the interacting scalar field
can be expand into momentum eigenmodes
\begin{eqnarray}\label{bf1}
\phi_x & = &
   \int\!\! \frac{d^3\vec{k}}{\sqrt{(2\pi)^3}}
   \left( a^\dagger_{k-}(t)\,\e^{-\mathrm{i}\vecs{k}\vecs{x}} +
          a_{k+}(t)\,        \e^{ \mathrm{i}\vecs{k}\vecs{x}}\right)
                \nonumber \\
\widetilde{\phi}_x & = &
   \int\!\! \frac{d^3\vec{k}}{\sqrt{(2\pi)^3}}
   \left( \widetilde{a}^\dagger_{k-}(t)\,\e^{\mathrm{i}\vecs{k}\vecs{x}} +
          \widetilde{a}_{k+}(t)        \,\e^{-\mathrm{i}\vecs{k}\vecs{x}}
                \right)
\;.\end{eqnarray}
$\vec{k}$ is the three-momentum of the modes, therefore in this notation
$a^\dagger_{k-}(t)$ creates a negatively charged excitation with momentum
$\vec{k}$, while $a_{k+}(t)$ annihilates a positive charge.
Henceforth the two different charges are distinguished
by an additional index $l=\pm$ whenever possible.

For the free case the commutation relations of the $a$-operators
at different times are
simple, while they are unknown for the interacting fields.
Note, that in the above expansion we have used a different
energy normalization factor than usual. While this might
be a little weird for free fields, it simplifies the calculations
for the interacting case.
\section{Free charged boson field}
The free complex scalar quantum fields are expand as specified
in eqn. (\ref{bf1}). The operators have commutation relations
\begin{eqnarray}\label{cn}
\left[a_{kl}(t),a^\dagger_{k^\prime l^\prime}(t)\right]&=&
  \frac{1}{\sqrt{2\omega_k}}\,
  \delta_{ll^\prime}\,
  \delta^3(\vec{k}-\vec{k}^\prime)\nonumber \\
\left[\widetilde{a}_{kl}(t),
  \widetilde{a}^\dagger_{k^\prime l^\prime}(t)\right]&=&
  \frac{1}{\sqrt{2\omega_k}}\,
  \delta_{ll^\prime}\,
  \delta^3(\vec{k}-\vec{k}^\prime)
\;,\end{eqnarray}
for $l=\pm$. All other commutators vanish, and $\omega$
is the free on-shell energy depending on momentum and
mass $m$ of the physical particles as
\begin{equation}\label{fosb}
\omega_k=\sqrt{\vec{k}^2+m^2}
\;.\end{equation}
The free physical states created and annihilated by the
above operators are stable. However,
they cannot be used as input for perturbative calculations,
since an infinitely small interaction gives them a finite
lifetime \cite{L88}.

The stable (albeit unobservable)
quasi-particle states are related to the physical states through
a Bogoliubov transformation. The hamiltonian of the system
is invariant under this transformation, if the system is in
global equilibrium. Hence, in this case the system has a global
Bogoliubov symmetry. The transformation can be written in matrix form as
\begin{eqnarray}\label{bbg1}
\left(\array{r} a_{kl}(t)\\
        \widetilde{a}^\dagger_{kl}(t) \endarray\right)&=&
({\cal B}(n_l(\vec{k})))^{-1}\,
\left(\array{r}\xi_{kl} \\ \widetilde{\xi}^\dagger_{kl}\endarray\right)
  \,\frac{1}{\sqrt{2\omega_k}}
  \,\e^{-\mathrm{i} \omega_kt} \nonumber \\
\left(\array{r}a^\dagger_{kl}(t) \\
       -\widetilde{a}_{kl}(t)\endarray\right)^T&=&
\left(\array{r}\xi^\ankh_{kl}\\
        -\widetilde{\xi}_{kl}\endarray\right)^T\,{\cal B}(n_l(\vec{k}))
  \frac{1}{\sqrt{2\omega_k}}\,
  \,\e^{\mathrm{i} \omega_kt}
\;.\end{eqnarray}
The $\xi$-operators create and annihilate stable momentum
eigenmodes, and the symbol $\xi^\ankh$ has been chosen to
indicate, that this operator is {\em not\/} the hermitean adjoint
of $\xi$. Rather, these operators are thermal conjugate to each other,
i.e., they annihilate the
thermal ground states according to
\begin{equation}\label{tsc}
\xi_{kl}\lds W^R \rdc = 0 \;\;,\;\;
\widetilde{\xi}_{kl}\lds W^R \rdc = 0 \;\;,\;\;
\ldc W^L\rds \xi^\ankh_{kl} = 0 \;\;,\;\;
\ldc W^L\rds \widetilde{\xi}^\ankh_{kl} = 0\;\;\;\forall\,\vec{k},l=\pm
\;\end{equation}
and have commutation relations
\begin{eqnarray}
\left[\xi_{kl},\xi^\ankh_{k^\prime l^\prime}\right]&=&
  \delta_{ll^\prime}\,
  \delta^3(\vec{k}-\vec{k}^\prime)\nonumber \\
\left[\widetilde{\xi}_{kl},\widetilde{\xi}^\ankh_{k^\prime l^\prime}\right]&=&
  \delta_{ll^\prime}\,
  \delta^3(\vec{k}-\vec{k}^\prime)
\;\end{eqnarray}
(all other commutators vanish).
Using these rules, any matrix element of field operators can be
expressed through the elements of the Bogoliubov matrix.

The Bogoliubov matrix can be parameterized in various ways,
leading to unitary inequivalent representations of the thermal
quasi-particles \cite{Ubook,hu92,X93}.
The most useful parameterization however
is obtained as linear function of a single parameter
$n$ that resembles the phase-space occupation
factor,
\begin{equation}\label{bdef}
 {\cal B}(n) =
\left(\array{lr}\left(1 + n\right) & -n\\
                 -1 & 1\endarray\right)
\;.\end{equation}
In general, the $n$-parameter can be different for any mode
of the system. The modes of the free scalar field are characterized by
their momentum, hence $n = n(\vec{k})$. The field is complex,
implying that the bosons carry a charge  and thus are not their own
anti-particles. Hence, for the two species of bosons,
the $n$-parameters can be different, $n_+\not\equiv n_-$.
If the charge they carry
is a conserved quantity, and the field is in thermal equilibrium,
we can chose Bose-Einstein functions as
\begin{equation}\label{nb0}
n_\pm(\vec{k}) = \frac{f_\pm(\vec{k})}{1-f_\pm(\vec{k})} =
        \frac{1}{\e^{\beta(\omega_k\mp\mu)}-1}
        \;\;,\;\;f^\pm(\vec{k}) = \e^{-\beta(\omega_k\mp\mu)}
\;.\end{equation}
Note, that in principle a generalized equilibrium state
with $\beta=\beta(\vec{k})$ could be used as well.

It is an easy task to calculate the propagator from the above
considerations, we refer to the existing literature or the next
section for its explicit form. More instructive for our purpose is the
case, where one uses a Bogoliubov transformation that mixes the
momenta as \cite{NUY92}
\begin{eqnarray}\label{bbg2}
\left({\array{r} a_{kl}(t)\\
         \widetilde{a}^\dagger_{kl}(t)\endarray}\right)&=
&
  \frac{1}{\sqrt{2\omega_k}}\,
\int\!\!d^3\vec{q}
  \left(\widetilde{\cal B}^{-1}_l(\vec{q},\vec{k})\right)^\star
  \,\left({\array{r}\xi_{ql}\\
              \widetilde{\xi}^\dagger_{ql}\endarray}\right)
  \,\e^{-\mathrm{i} \omega_qt} \nonumber \\
\left({\array{r}a^\dagger_{kl}(t)\\
        -\widetilde{a}_{kl}(t)\endarray}\right)^T&=
&
  \frac{1}{\sqrt{2\omega_k}}\,
\int\!\!d^3\vec{q}
  \left({\array{r}\xi^\ankh_{ql}\\
           -\widetilde{\xi}_{ql}\endarray}\right)^T\,
  \widetilde{\cal B}_l(\vec{q},\vec{k})
  \,\e^{\mathrm{i} \omega_qt}
\;.\end{eqnarray}
This momentum mixing occurs, when one makes
the Bogoliubov symmetry mentioned before
a {\em local\/} symmetry. Such a process is called
the {\em gauging\/} of the symmetry, and in a quite natural way
introduces a couping to external parameters into the model \cite{h90ber}.
The generalized Bogoliubov matrix is expressed as
\begin{equation}\label{gb}
\widetilde{\cal B}_l(\vec{q},\vec{k}) = \left( { \array{lr}
   \left(\delta^3(\vec{q}-\vec{k}) + N_l(\vec{q},\vec{k})\right)
            \;\;\;& -N_l(\vec{q},\vec{k}) \\
   -\delta^3(\vec{q}-\vec{k})     & \delta^3(\vec{q}-\vec{k})
   \endarray} \right)
\;.\end{equation}
It obeys the relations
\begin{eqnarray}
\left(\widetilde{\cal B}^{-1}_l(\vec{q},\vec{k})\right)^\star\,
\widetilde{\cal B}_l(\vec{q},\vec{k}^\prime)
 &=& \delta^3(\vec{k}-\vec{q})\,
     \delta^3(\vec{q}-\vec{k}^\prime)\,\mbox{diag}(1,1) \nonumber \\
 &+& T_0\tau_3\,N_l(\vec{k},\vec{k}^\prime)
 \left(\delta^3(\vec{k}-\vec{q})-\delta^3(\vec{q}-\vec{k}^\prime\right)
 \nonumber \\
\int\!\!\!d^3\vec{q}\,
\left(\widetilde{\cal B}^{-1}_l(\vec{q},\vec{k})\right)^\star\,
\widetilde{\cal B}_l(\vec{q},\vec{k}^\prime)
&=&\delta^3(\vec{k}-\vec{k}^\prime)\,\mbox{diag}(1,1)
\;,\end{eqnarray}
with the Pauli matrix
$\tau_3=\mathrm{diag}(1,-1)$ and
\begin{equation}
T_0=\left({\array{lr} 1 & 1\\ 1 & 1 \endarray}\right)\,
\;.\end{equation}
The occupation number density parameter appearing here is the Fourier
transform of a space-local quantity
\begin{equation}\label{nloc0}
N_l(\vec{q},\vec{k}) =
  \frac{1}{(2\pi)^3}\;\int\!\!d^3\vec{z}\,
   \e^{-\mathrm{i}(\vecs{q}-\vecs{k})\vecs{z}}
  \,n_l((\vec{q}+\vec{k})/2,\vec{z})
\;,\end{equation}
and we specify that the $n_l(\vec{k},\vec{z})$ are real functions.
This implies, that $N^\star_l(\vec{q},\vec{k})=N_l(\vec{k},\vec{q})$.
Moreover, the generalized Bogoliubov matrix from eqn. (\ref{gb})
is related to the one from (\ref{bdef}) as
\begin{equation}
\widetilde{\cal B}_l(\vec{q},\vec{k})=
  \frac{1}{(2\pi)^3}\;\int\!\!d^3\vec{z}\,
   \e^{-\mathrm{i}(\vecs{q}-\vecs{k})\vecs{z}}\;
  \,{\cal B}(n_l((\vec{q}+\vec{k})/2,\vec{z}))
\;.\end{equation}
As one can check within a few lines, the above transformation
preserves the canonical commutation relations. However,
the time evolution is complicated by the simple fact that
the phase factors in eqn. (\ref{bbg2}) carry the momentum index of the
$\xi$-operators. Indeed, one obtains for the time derivative
of the $a$-operators
\begin{eqnarray}\label{tidz}
 \mathrm{i}\frac{\partial}{\partial t}\,
 \left({\array{r} a_{kl}(t)\\
         \widetilde{a}^\dagger_{kl}(t)\endarray}\right)&=&
 \omega_k\,\left({\array{r} a_{kl}(t)\\
         \widetilde{a}^\dagger_{kl}(t)\endarray}\right) \nonumber \\
& +&\int\!\!d^3\vec{q}
  \,\sqrt{\frac{\omega_q}{\omega_k}}
  \;N_l(\vec{k},\vec{q})\,\left(\omega_k-\omega_q\right)\,T_0\tau_3\,
  \left({\array{r}a_{ql}(t)\\
           \widetilde{a}^\dagger_{ql}(t)\endarray}\right)
\;.\end{eqnarray}
A similar relation is obtained for the adjoint operators. Note,
that the additional term in this time derivative vanishes, if
$N(\vec{q},\vec{k})$ is proportional to
$\delta^3(\vec{q}-\vec{k})$.  This is the case, when
$n(\vec{k},\vec{z})$ does not depend on the space coordinate
$\vec{z}$, i.e., when the system
is translationally invariant.

We can now ask, what kind of Hamilton operator $\widehat{H}$ would give a
time derivative according to eqn. (\ref{tidz}) if used in the
Heisenberg equation
\begin{equation}\label{heis}
 \mathrm{i}\frac{\partial}{\partial t}\,
 \left({\array{r} a_{kl}(t)\\
         \widetilde{a}^\dagger_{kl}(t)\endarray}\right)\;=\;
 \left[\left({\array{r} a_{kl}(t)\\
         \widetilde{a}^\dagger_{kl}(t)\endarray}\right),\widehat{H}\right]
\;.\end{equation}
Clearly, such a hamiltonian has two parts: one ''bare'' giving the
unperturbed time evolution
\begin{equation}\label{hnn}
  \widehat{H}_0 = \sum\limits_{l=\pm}
\int\!\!d^3\vec{k}\,2\omega_k^2\,
        \left({\array{r} a^\dagger_{kl}(t)\\
         -\widetilde{a}_{kl}(t)\endarray}\right)^T\,
        \left({\array{r} a_{kl}(t)\\
         \widetilde{a}^\dagger_{kl}(t)\endarray}\right)
\;,\end{equation}
and a part that vanishes for homogeneous systems:
\begin{equation}\label{hen}
\widehat{H}_1 = \sum\limits_{l=\pm}
\int\!\!d^3\vec{k}\,d^3\vec{q}\,
        \left({\array{r} a^\dagger_{kl}(t)\\
         -\widetilde{a}_{kl}(t)\endarray}\right)^T\,T_0\tau_3\,
        {\cal H}_l(\vec{k},\vec{q})\,
        \left({\array{r} a_{ql}(t)\\
         \widetilde{a}^\dagger_{ql}(t)\endarray}\right)\,
\;,\end{equation}
with
\begin{equation}
{\cal H}_l(\vec{k},\vec{q})
        = N_l(\vec{k},\vec{q})\,
          2\sqrt{\omega_k\omega_q}\,
          \left(\omega_k-\omega_q\right)
\;.\end{equation}
Note, that the dimension of the numerical factors in
(\ref{hnn}) and (\ref{hen}) is mass$^2$, due to the
energy normalization factors contained in the
$a$-operators, cf. eqn. (\ref{cn}).
The further discussion of this modified hamiltonian we
postpone until the above expression has been generalized
to interacting systems.
\section{Interacting fields in global equilibrium}
The procedure to derive thermal Green's functions for interacting
fields has been outlined in refs. \cite{hu92,hu92a}, and thus will
not be repeated here in full detail. It is based on the expansion
of the interacting field into modes with definite energy and momentum
according to ref. \cite{L88}. Their superposition with a positive
weight function then gives the fully interacting field.

Such an expansion is necessary, because at finite density and
temperature the irreducible representations of the space-time
symmetry group are characterized by two continuous parameters \cite{BS75}.
Since these fields contain the full interaction of the system,
the commutation relations of the fields with definite energy and momentum
are not known in general. However, for the task of calculating
two-point functions or bilinear expectation values, one needs
to know only the expectation value of such commutators -- and
this information can be absorbed into the weight function.

In other words, for the calculation of bilinear expectation values
of interacting fields it is
sufficient to consider the modes of definite energy and momentum
as generalized free fields \cite{L88}. The full information about the
single-particle spectrum of the theory is contained in the
weight functions.

The final step in deriving full propagators at finite temperature
then consists in applying the thermal quasi-particle picture to
each of these modes. We thus write for the operators of the field
expansion
\begin{eqnarray}\label{bbg3}
\left({\array{r} a_{k\pm}(t)\\
          \widetilde{a}^\dagger_{k\pm}(t)\endarray}\right)\;=
& \int\limits_0^\infty\!\!dE
  \;\rho^{1/2}_\pm(E,\vec{k})\,
  \left({\cal B}(n_\pm(E,\vec{k}))\right)^{-1}
  \,\left({\array{r}\xi_{Ek\pm}\\
              \widetilde{\xi}^\dagger_{Ek\pm}\endarray}\right)
  \,\e^{-\mathrm{i} Et} \nonumber \\
\left({\array{r}a^\dagger_{k\pm}(t)\\
         -\widetilde{a}_{k\pm}(t)\endarray}\right)^T\;=
& \int\limits_0^\infty\!\!dE
  \;\rho^{1/2}_\pm(E,\vec{k})\,
  \left({\array{r}\xi^\ankh_{Ek\pm}\\
           -\widetilde{\xi}_{Ek\pm}\endarray}\right)^T\,
  {\cal B}(n_\pm(E,\vec{k}))
  \,\e^{\mathrm{i} Et}
\;.\end{eqnarray}
This Bogoliubov transformation for
interacting systems defines stable, albeit non-observable
quasi-particles \cite{YUNA92,hu92,hu92a}, created and
annihilated by the $\xi$-operators.
The $\xi$-operators have different commutation relations,
cf. ref. \cite{L88}
\begin{eqnarray}\label{difc}
\left[\xi_{Ekl},\xi^\ankh_{E^\prime k^\prime l^\prime}\right]&=&
  \delta_{ll^\prime}\,
  \delta(E-E^\prime)\,
\delta^3(\vec{k}-\vec{k}^\prime)\nonumber \\
\left[\widetilde{\xi}_{Ekl},
\widetilde{\xi}^\ankh_{E^\prime k^\prime l^\prime}
  \right]&=&
  \delta_{ll^\prime}\,
  \delta(E-E^\prime)\,
  \delta^3(\vec{k}-\vec{k}^\prime)
\;\end{eqnarray}
(all other commutators vanish), but the thermal state conditions remain
similar to (\ref{tsc})
\begin{equation}\label{tscc}
\xi_{Ekl}\lds W^R \rdc = 0 \;\;,\;\;
\widetilde{\xi}_{Ekl}\lds W^R \rdc = 0 \;\;,\;\;
\ldc W^L\rds \xi^\ankh_{Ekl} = 0 \;\;,\;\;
\ldc W^L\rds \widetilde{\xi}^\ankh_{Ekl} = 0\;\;\;\forall\,E,\vec{k},l=\pm
\;. \end{equation}
With these rules, all bilinear expectation values can be
calculated. Matrix elements of higher powers of interacting field operators
however are obtained only approximately, they have a perturbative
expansion in terms of the weight functions $\rho_l(E,\vec{k})$
\cite{L88}.

$n$ is a Bose-Einstein distribution function, which for
special case of a global equilibrium depends only on
the energy of the modes:
\begin{equation}\label{nb}
n_\pm(E,\vec{k}) = \frac{f_\pm(E)}{1-f_\pm(E)} =
\frac{1}{\e^{\beta (E\mp\mu)}-1}
        \;\;,\;\;f_\pm(E) = \e^{-\beta( E\mp\mu) }
\;.\end{equation}
In our notation, the weight functions
$\rho_l(E,\vec{k})$ have support only for positive energies,
and their integrals are
\begin{eqnarray}\label{norm}
\int\limits_0^\infty\!\!dE\,E\, \rho_l(E,\vec{k})& =&\frac{1}{2}
\nonumber \\
\int\limits_0^\infty\!\!dE \, \rho_l(E,\vec{k}) &=&Z_{kl}
\;.\end{eqnarray}
Note, that the existence of this spectral
decomposition is only guaranteed in case the system is space-time
translation invariant, i.e., if it is in a thermal equilibrium
state. It follows from these definitions, that
\begin{eqnarray}\label{co}
\left[a_{kl}(t),a^\dagger_{k^\prime l^\prime}(t)\right]&=&
  Z_{kl}\,
  \delta_{ll^\prime}\,
  \delta^3(\vec{k}-\vec{k}^\prime)\nonumber \\
\left[\widetilde{a}_{kl}(t),
  \widetilde{a}^\dagger_{k^\prime l^\prime}(t)\right]&=&
  Z_{kl}\,
  \delta_{ll^\prime}\,
  \delta^3(\vec{k}-\vec{k}^\prime)
\; \end{eqnarray}
in generalization of eqn. (\ref{cn}). Collecting these definitions,
the full bosonic two-point Green's function, or propagator,
at finite temperature is derived as  \cite{hu92,hu92a}
\begin{eqnarray}
&&D^{(ab)}(k_0,\vec{k})= \nonumber \\
&&\int\limits_0^\infty\!\!dE\,
 \rho_+(E,\vec{k})\;
  ({\cal B}(n_+(E,\vec{k})))^{-1}
  \left(\!{\array{ll}
         {\displaystyle \frac{1}{k_0-E+\mathrm{i}\epsilon}} & \\
    &    {\displaystyle \frac{1}{k_0-E-\mathrm{i}\epsilon}}
\endarray} \right)\;
         {\cal B}(n_+(E,\vec{k}))\,\tau_3\nonumber \\
&-&\int\limits_0^{\infty}\!\!dE\,
 \rho_-(E,\vec{k})\;
  \tau_3\,{\cal B}^T(n_-(E,\vec{k}))
  \left(\!{\array{ll}
         {\displaystyle \frac{1}{k_0+E-\mathrm{i}\epsilon}} & \\
    &    {\displaystyle \frac{1}{k_0+E+\mathrm{i}\epsilon}}
\endarray} \right)\;
         ({\cal B}^T(n_-(E,\vec{k})))^{-1}
\;.\end{eqnarray}
In the above equation, the terms
propagating particle and anti-particle states have been kept
separately. Eqn. (\ref{nb}) can be continued to negative
energy arguments, and
\begin{equation}
n_-(E,\vec{k}) = - (1+n_+(-E,-\vec{k}))
\,.\end{equation}
Setting $n_+(E,\vec{k})\equiv n(E)$ in the following,
the above propagator simplifies to
\begin{eqnarray}\label{dbk}
D^{(ab)}(k_0,\vec{k})&& = \int\limits_{-\infty}^\infty\!\!dE\,
 {\cal A}(E,\vec{k})\;\times\nonumber\\
  &&({\cal B}(n(E)))^{-1}
  \left(\!{\array{ll}
         {\displaystyle \frac{1}{k_0-E+\mathrm{i}\epsilon}} & \\
    &    {\displaystyle \frac{1}{k_0-E-\mathrm{i}\epsilon}}
\endarray} \right)\;
         {\cal B}(n(E))\,\tau_3
\;.\end{eqnarray}
${\cal A}(E,\vec{k})$ is the spectral function of the boson field,
\begin{equation}\label{spec}
{\cal A}(E,\vec{k}) = \rho_+(E,\vec{k})\Theta(E) -
                        \rho_-(-E,-\vec{k})\Theta(-E)
\;,\end{equation}
and the limit of free particles with mass  $m$ is recovered when
\begin{equation}\label{fb}
{\cal A}(E,\vec{k}) \longrightarrow
\mbox{sign}(E)\,
  \delta(E^2 -\vec{k}^2 -m^2)
=\mbox{sign}(E)\,
  \delta(E^2 -\omega_k^2)
\;.\end{equation}
\section{Interacting fields with local Bogoliubov symmetry}
As was stated above, the existence of a spectral decomposition
is bound to the equilibrium property of the system, i.e., to
its space-time translation invariance \cite{Ubook}.
We can expect however,
that close to equilibrium the field properties do not change very
much. Thus, the considerations from the second section of the present
paper can be carried over to the interacting case: We want to study
the quantum system under the influence of small gradients in the
distribution function $n$, but with the approximation of
a space-time independent spectral weight function $\rho_l(E,\vec{k})$.
In generalization of eqn. (\ref{bbg2}):
\begin{eqnarray}\label{bbg4}
\left({\array{r} a_{kl}(t)\\
          \widetilde{a}^\dagger_{kl}(t)\endarray}\right)\;=
& \int\limits_0^\infty\!\!dE\,\int\!\!d^3\vec{q}
  \;\rho^{1/2}_l(E,(\vec{q}+\vec{k})/2)\,
  \left(\widetilde{\cal B}^{-1}_l(E,\vec{q},\vec{k})\right)^\star
  \,\left({\array{r}\xi_{Eql}\\
              \widetilde{\xi}^\dagger_{Eql}\endarray}\right)
  \,\e^{-\mathrm{i} Et} \nonumber \\
\left({\array{r}a^\dagger_{kl}(t)\\
         -\widetilde{a}_{kl}(t)\endarray}\right)^T\;=
& \int\limits_0^\infty\!\!dE\,\int\!\!d^3\vec{q}
  \;\rho^{1/2}_l(E,(\vec{q}+\vec{k})/2)\,
  \left({\array{r}\xi^\ankh_{Eql}\\
           -\widetilde{\xi}_{Eql}\endarray}\right)^T\,
  \widetilde{\cal B}_l(E,\vec{q},\vec{k})
  \,\e^{\mathrm{i} Et}
\;.\end{eqnarray}
The transformation matrix here has the same functional form
as in eqn. (\ref{gb}), but now its occupation number density
parameter also depends on the energy variable
\begin{equation}\label{nloc}
N_l(E,\vec{q},\vec{k}) = \frac{1}{(2\pi)^3}\;
        \int\!\!d^3\vec{z}\,\e^{-\mathrm{i}
  (\vecs{q}-\vecs{k})\vecs{z} }\,
  n_l(E,(\vec{q}+\vec{k})/2,\vec{z})
\;.\end{equation}
The propagator matrix according to (\ref{pd})
is calculated as
\begin{eqnarray}\label{fbtc}
&&D^{(ab)}(t,\vec{x};t^\prime,\vec{x}^\prime)=
-\mathrm{i}\int\limits_0^\infty\!\!dE\,\int\!\!
     \frac{d^3\vec{k}\,d^3\vec{q}\,d^3\vec{k}^\prime}{
           (2\pi)^3}\,\rho^{1/2}_+(E,(\vec{k}+\vec{q})/2)
                    \,\rho^{1/2}_+(E,(\vec{q}+\vec{k}^\prime)/2)
\times \nonumber\\
&&  (\widetilde{\cal B}^{-1}_+(E,\vec{q},\vec{k}))^\star\;
    \left(\!{\array{ll} \Theta(t-t^\prime) & \\
               &   -\Theta(t^\prime-t) \endarray}\right)
     \widetilde{\cal B}_+(E,\vec{q},\vec{k}^\prime)\tau_3
         \,\e^{-\mathrm{i} E(t-t^\prime)}
         \e^{\mathrm{i}\vecs{k}\vecs{x}-
             \mathrm{i}\vecs{k}^\prime\vecs{x}^\prime}
         \nonumber \\
&&\phantom{D^{(ab)}(t,\vec{x};t^\prime,\vec{x}^\prime)=}
-\mathrm{i}\int\limits_0^\infty\!\!dE\,\int\!\!
     \frac{d^3\vec{k}\,d^3\vec{q}\,d^3\vec{k}^\prime}{
           (2\pi)^3}\,\rho^{1/2}_-(E,(\vec{k}+\vec{q})/2)
                    \,\rho^{1/2}_-(E,(\vec{q}+\vec{k}^\prime)/2)
\times \nonumber\\
&&  \tau_3\widetilde{\cal B}^T_-(E,\vec{q},\vec{k})\;
    \left(\!{\array{ll} \Theta(t^\prime-t) & \\
               &   -\Theta(t-t^\prime) \endarray}\right)
   \left(\widetilde{\cal B}^T_-(E,
   \vec{q},\vec{k}^\prime)\right)^{\star-1}
   \,\e^{\mathrm{i} E(t-t^\prime)}
         \e^{-\mathrm{i}\vecs{k}\vecs{x}+
              \mathrm{i}\vecs{k}^\prime\vecs{x}^\prime}
\;.\end{eqnarray}
Similar to the equilibrium case and to the
time-dependent case one can perform the energy integration
because of the linearity of this expression in $n$. This
allows to write the full propagator in terms of Bogoliubov
matrices, which depend on space-local distribution functions.
The principles of this method are outlined in ref. \cite{hu92},
and will not be discussed in the present paper.

As the last point in considering the above full Green function,
consider the imaginary part of the retarded propagator obtained as
\begin{eqnarray}
\mbox{Im}\left(D^{R}(t,\vec{x};t^\prime,\vec{x}^\prime)\right) &=&
\mbox{Im}\left(D^{(11)}(t,\vec{x};t^\prime,\vec{x}^\prime) -
                 D^{(12)}(t,\vec{x};t^\prime,\vec{x}^\prime)\right)
 \nonumber \\
&=&-\pi \int\!\!\frac{dEd^3\vec{k}}{(2\pi)^4}\,
\e^{-\mathrm{i}E(t-t^\prime)+
      \mathrm{i}\vecs{k}(\vecs{x}-\vecs{x}^\prime)}\,
\left(\rho_+(E,\vec{k}) -  \rho_-(-E,-\vec{k})\right)
\;.\end{eqnarray}
The retarded and advanced propagator hence only depend on the coordinate
differences, and their imaginary part still gives
the spectral function from eqn. (\ref{spec}). This is the only approximation
made so far, but it can be shown easily by solving the matrix-valued
Schwinger-Dyson equation in coordinate representation, that corrections
to the above spectral functions are {\em of second order in the gradients}.

In generalization of eqn. (\ref{tidz}), one thus obtains for
the time derivative
\begin{eqnarray}\label{tide}
 \mathrm{i}\frac{\partial}{\partial t}\,
 \left({\array{r} a_{kl}(t)\\
         \widetilde{a}^\dagger_{kl}(t)\endarray}\right)&=&
 \Omega_{kl}\,\left({\array{r} a_{kl}(t)\\
         \widetilde{a}^\dagger_{kl}(t)\endarray}\right) \nonumber \\
& +&\int\limits_0^\infty\!\!dE\int\!\!d^3\vec{q}
  \,\frac{\rho^{1/2}_l(E,(\vec{k}+\vec{q})/2)}{Z_{ql}}
  \;N_l(E,\vec{k},\vec{q})\,\times\nonumber \\
&&  \left(E\,\rho^{1/2}_l(E,\vec{k})-
        E\,\rho^{1/2}_l(E,\vec{q})\right)
   \,T_0\tau_3\,
  \left({\array{r}a_{ql}(t)\\
           \widetilde{a}^\dagger_{ql}(t)\endarray}\right)
\;.\end{eqnarray}
Here, the energy $\Omega$ is defined as
\begin{equation}
\Omega_{kl}  = \frac{1}{Z_{kl}}\,\int\limits_0^\infty\!\!dE\,E\,
        \rho_l(E,\vec{k})\;=\;\frac{1}{2Z_{kl}}
\;,\end{equation}
and $Z_{kl}$ as in eqn. (\ref{norm}). Obviously,
as for the free field, the additional term
in this time evolution vanishes for translationally invariant
$n$, i.e., when $N(E,\vec{k},\vec{q})$ is proportional to
$\delta^3(\vec{k}-\vec{q})$. It has a generator $\widehat{H}_1$,
which has the same functional form as in eqn. (\ref{hen}),
but the kernel reads
\begin{equation}
{\cal H}_l(\vec{k},\vec{q})
        =\int\limits_0^\infty\!\!dE
  \,\left\{\frac{\rho^{1/2}_l(E,(\vec{k}+\vec{q})/2)}{
                 Z_{kl}\,Z_{ql}}\right\}
  \,N_l(E,\vec{k},\vec{q})\,
  \left(E\,\rho^{1/2}_l(E,\vec{k})-
        E\,\rho^{1/2}_l(E,\vec{q})\right)
\;.\end{equation}
The perturbation of the system has
a very special form given in eqn. (\ref{hen}). If it is expressed
in terms of the $\xi$-operators, one may use
\begin{equation}
\widetilde{\cal B}(E,\vec{q},\vec{q}^\prime)\,
    \left({\array{lr} 1 & 1\\ 1 & 1 \endarray}\right)\,
  (\widetilde{\cal B}^{-1}(E,\vec{k},\vec{k}^\prime))^\star
  = \left({\array{lr} 0 &-1\\ 0 & 0 \endarray}\right)\,
    \delta^3(\vec{q}-\vec{q}^\prime)\,
    \delta^3(\vec{k}-\vec{k}^\prime)\,
\;\end{equation}
to see that only the combination $\xi^\ankh\widetilde{\xi}^\ankh$
survives:
\begin{equation}\label{hens}
\widehat{H}_1 = \sum\limits_{l=\pm}
\int\limits_0^\infty\!\!dE\,dE^\prime
\int\!\!d^3\vec{k}\,d^3\vec{k}^\prime\,
        \rho^{1/2}_l(E,\vec{k})\,\rho^{1/2}_l(E^\prime,\vec{k}^\prime)
        \;{\cal H}_l(\vec{k},\vec{k}^\prime)\;
             \xi^\ankh_{Ekl}
             \widetilde{\xi}^\ankh_{E^\prime k^\prime l}
\;.\end{equation}
This simple expression for the gradient part of the hamiltonian can
be inserted into eqn. (\ref{td}). Note, that the total hamiltonian
is diagonal in the $\xi$-operators, but not if expressed in terms of the
physical particle operators $a$, $a^\dagger$.
\section{Transport properties of an interacting scalar field}
Observables are expressed as functionals of the interacting fields
-- and therefore as functionals of the operators $a$, $a^\dagger$.
Hence, although the state we consider is stationary in terms of
the basis defined by the $\xi$-operators, momentum mixing as
introduced above will result in a non-trivial time evolution
of {\em physical\/} quantities, according to the Heisenberg
equation (\ref{heis}). Thus, equation (\ref{kubo}) can be used
to calculate relevant contributions to system parameters due to
the spatial dependence of $n(E,\vec{k},\vec{x})$.

Two quantities we are interested in are the
conserved particle current operator of the complex scalar field
and its expectation value,
\begin{equation}
\widehat{\vec{j}}(t,\vec{x}) = \mathrm{i}\,
   \left(\phi^\star_x\nabla\phi_x - \phi_x\nabla\phi^\star_x
   \right)\;\;\;\;
   \vec{j}(t,\vec{x})=
    \delta\Av{\widehat{\vec{j}}(t,\vec{x})}
\;\end{equation}
and the conserved energy current
\begin{equation}
\widehat{\vec{E}}(t,\vec{x}) =
   \left(\partial_t\phi^\star_x\nabla\phi_x +
   \partial_t\phi_x\nabla\phi^\star_x
   \right)\;\;\;\;
   \vec{E}(t,\vec{x})=
    \delta\Av{\widehat{\vec{E}}(t,\vec{x})}
\;.\end{equation}
Inserting these into eqn. (\ref{td}) then yields an expression
for the commutator, which contains {\em four\/} field
operators instead of {\it two\/}. The expectation value of this
commutator therefore cannot be expressed completely by the interacting
two-point functions we have calculated. For these, it had been
explicitly assumed above, that the fields can be written
as generalized free fields.

However, the four-field expectation value has a diagrammatic expansion
in terms of the propagators we derived - and the lowest order
term of this expansion already contains the full single-particle
spectrum. We expect, that this lowest order contributes
the dominant part of the system's response.

Using the commutation rules for the generalized free fields,
e.g. the Bogoliubov transformation and (\ref{difc}), therefore
amounts to the neglection of higher correlations. These however
can be incorporated through the appropriate Feynman rules
\cite{L88,H90}.

If we insert the expansion of the fields according to (\ref{bf1}),
all kinds of binary products of the $a$-operators appear
in these operator expressions.
However, in the present formulation, out
of 4 $\times$ 8 combinations of $a$-operators contained in a naive
calculation of the commutator (\ref{kubo}), only two survive
due to the special form (\ref{hens}) of $\widehat{H}_1$. These are
\begin{eqnarray}
&&
\Av{\left[a_{kl}^\dagger(t-\tau)\,a_{k^\prime l}(t-\tau),
          \xi^\dagger_{E^\prime q^\prime \,l}\,
          \widetilde{\xi}^\dagger_{Eq\,l}\right]} =
          \nonumber \\
          &&\;\;\;\;\;\;\;\;\;
   = \rho_l^{1/2}(E,\vec{q})\,\rho_l^{1/2}(E^\prime,\vec{q}^\prime)
   \,\delta^3(\vec{k}-\vec{q})\,\delta^3(\vec{k}^\prime-\vec{q}^\prime)
   \,\e^{\mathrm{i}(E-E^\prime)(t-\tau)}\nonumber \\
&&
\Av{\left[\partial_ta_{kl}^\dagger(t-\tau)\,a_{k^\prime l}(t-\tau),
          \xi^\dagger_{E^\prime q^\prime \,l}\,
          \widetilde{\xi}^\dagger_{Eq\,l}\right]} =
          \nonumber \\
          &&\;\;\;\;\;\;\;\;\;
   = \mathrm{i}E\,
   \rho_l^{1/2}(E,\vec{q})\,\rho_l^{1/2}(E^\prime,\vec{q}^\prime)
   \,\delta^3(\vec{k}-\vec{q})\,\delta^3(\vec{k}^\prime-\vec{q}^\prime)
   \,\e^{\mathrm{i}(E-E^\prime)(t-\tau)}
\;\end{eqnarray}
for $l=\pm$, i.e., both charge species.

We can then assemble the results for the currents,
by performing the time integration as if $t\rightarrow\infty$
and $t_0\rightarrow-\infty$. Alternatively, one could first
process the momentum integrations and then pick out the real part of the
current. One obtains
\begin{equation}
\vec{j}_l(t,\vec{x}) =  -2\pi \mathrm{i}\,
\int\limits_0^\infty\!\!dE\,
\int\!\!\frac{d^3\vec{k}\,d^3\vec{k}^\prime}{(2\pi)^3}\,
        \rho_l(E,\vec{k})\,\rho_l(E,\vec{k}^\prime)
        \,\e^{\mathrm{i}(\vecs{k}-\vecs{k}^\prime)\vecs{x}}
        \,\left(\vec{k}+\vec{k}^\prime\right)
        \,{\cal H}_l(\vec{k},\vec{k}^\prime)
\;\end{equation}
for each species. The energy current is formally quite similar,
\begin{equation}
\vec{E}_l(t,\vec{x}) = -2\pi\mathrm{i}\,
\int\limits_0^\infty\!\!dE\,
\int\!\!\frac{d^3\vec{k}\,d^3\vec{k}^\prime}{(2\pi)^3}\,
        \rho_l(E,\vec{k})\,\rho_l(E,\vec{k}^\prime)
        \,\e^{\mathrm{i}(\vecs{k}-\vecs{k}^\prime)\vecs{x}}
        \,E\left(\vec{k}+\vec{k}^\prime\right)
        \,{\cal H}_l(\vec{k},\vec{k}^\prime)
\;\end{equation}
and the total currents are for both cases
\begin{eqnarray}
\vec{j}(t,\vec{x}) &=&
\vec{j}_+(t,\vec{x})-\vec{j}_-(t,\vec{x}) \nonumber \\
\vec{E}(t,\vec{x})&=&
\vec{E}_+(t,\vec{x})+\vec{E}_-(t,\vec{x})
\;.\end{eqnarray}
In principle the above expressions can be calculated, when
the spectral functions and the space-dependence of $n$ are
given. However, it is more instructive to perform a gradient
expansion for the occupation number density $n$. For this we use
the abbreviation $\vec{Q}=(\vec{k}+\vec{k}^\prime)/2$, and
expand all momenta around this value, e.g.
\begin{equation}\label{gex}
  \rho^{1/2}_l(E,\vec{Q})
  \left(E\,\rho^{1/2}_l(E,\vec{k})-
        E\,\rho^{1/2}_l(E,\vec{k}^\prime)\right)
  \;\approx\; (\vec{k}-\vec{k}^\prime)\,\frac{E}{2}\,
  \frac{\partial}{\partial \vec{Q}}
  \rho_l(E,\vec{Q})
\:.\end{equation}
The factor $(\vec{k}-\vec{k}^\prime)$ can be taken out of the integrations,
as $-\mathrm{i}\partial/\partial\vec{x}\equiv-\mathrm{i}\nabla_x$.
Other contributions are of zeroth and second order in this derivative, e.g.
\begin{equation}
 \rho_l(E,\vec{k})\,\rho_l(E,\vec{k}^\prime)
=\rho_l^2(E,\vec{Q}) + {\cal O}\left((\vec{k}-\vec{k}^\prime)^2\right)
\;.\end{equation}
Therefore, the $i$-th vector component of the $l$-charged currents
generated by the inhomogeneity of the system is
\begin{eqnarray}\label{cur}
\vec{j}^{(i)}_l(t,\vec{x})&=\;
   2\pi\,\int\!\!\frac{d^3\vec{Q}}{(2\pi)^3}\,
   \frac{\vec{Q}^{(i)}}{2\,Z^2_{Ql}}&
   \int\!\!dE\,\left(\rho_l(E,\vec{Q})\right)^2
   \times\nonumber \\
 &&\int\!\!dE^\prime\,E^\prime\,\left\{
   \frac{\partial n_l(E^\prime,\vec{Q},\vec{x})}{\partial \vec{x}^{(j)}}
   \frac{\partial \rho_l(E^\prime,\vec{Q})}{\partial \vec{Q}^{(j)}}
   \right\} \nonumber \\
\vec{E}^{(i)}_l(t,\vec{x})&=\;
   2\pi\,\int\!\!\frac{d^3\vec{Q}}{(2\pi)^3}\,
   \frac{\vec{Q}^{(i)}}{2\,Z^2_{Ql}}&
   \int\!\!dE\,E\,\left(\rho_l(E,\vec{Q})\right)^2
   \times\nonumber \\
 &&\int\!\!dE^\prime\,E^\prime\,\left\{
   \frac{\partial n_l(E^\prime,\vec{Q},\vec{x})}{\partial \vec{x}^{(j)}}
   \frac{\partial \rho_l(E^\prime,\vec{Q})}{\partial \vec{Q}^{(j)}}
   \right\}
\;.\end{eqnarray}
Before this is evaluated further, note that the expression
in the curly brackets is nothing but the first term of a full
gradient expansion of the product of $n$ and $\rho$, i.e.,
the Poisson bracket. It is therefore clear, that the above expression
is useful also for situations, where the space-dependence of
$\rho$ cannot be neglected: the Poisson bracket then also has
a contribution with a space-derivative acting on $\rho$ and a
momentum derivative acting on $n$.
The above expression is also interesting in view of the
original equation (\ref{td}), since the Poisson bracket
is the analogon of the commutator, i.e., it contributes
a factor $\hbar$ to the the current.

To proceed, some remarks on the derivatives occuring in the
Poisson bracket are necessary. It was assumed,
that the system is in a local equilibrium state,
close to global equilibrium such that the gradients are small
and currents are only due to these gradients. This implies, that
the spectral function depends only on $Q=\left|\vec{Q}\right|$,
and thus the momentum derivative gives
\begin{equation}
\frac{\partial \rho_l(E^\prime,\vec{Q})}{\partial \vec{Q}^{(j)}}
=\frac{\vec{Q}^{(j)}}{Q}\,\frac{\partial \rho_l(E^\prime,Q)}{\partial Q}
\;\end{equation}
The currents from eqn. (\ref{cur})
thus are symmetric in the indices $i$ and $j$.

Using as local equilibrium
distribution function instead of eqn. (\ref{nb}) the expression
\begin{equation}\label{nb2}
n_\pm(E,\vec{Q},\vec{x}) = \frac{f_\pm(E,\vec{x})}{1-f_\pm(E,\vec{x})} =
\frac{1}{\e^{\beta(\vecs{x}) (E\mp\mu(\vecs{x}))}-1}
        \;\;,\;\;f_\pm(E,\vec{x}) = \e^{-\beta(\vecs{x})(
    E\mp\mu(\vecs{x}))}
\;,\end{equation}
the derivative with respect to $\vec{x}$ has two components,
\begin{equation}\label{qdi}
\frac{\partial}{\partial \vec{x}^{(j)}}
n_\pm(E,\vec{Q},\vec{x}) = \frac{1}{4\sinh^2(\beta(E\mp\mu)/2)}
\,\left( \frac{E\mp\mu}{T}\,\frac{\nabla^{(j)}_x T}{T}\pm\frac{1}{T}
  \,\nabla^{(j)}_x\mu\right)
\;.\end{equation}
This derivative however has to be taken with care:
In eqns. (\ref{cur}),
the $\vec{x}$-derivative was shifted through the integrals as if they
were all properly defined. This requires, in the absence of
a Bose condensate, that the spectral function vanishes linearly
at $E=\mu$. A discussion of this requirement will be attempted
in a separate publication, here we will take it for granted.

However, this violates the assumption of a spectral function
independent of the gradients, since the $\vec{x}$-derivative
then also acts on $\rho(E,\vec{k})$. To avoid this complication,
we define the integral over the distribution
function as a principal value integral, i.e., it
is only defined in the sense of first calculating
\begin{equation}\label{npr}
I_{Ql}=\int\limits_0^\infty\!\!dE\,\rho(E,\vec{Q})\,n(E,\vec{Q},
\vec{x})=
\lim_{\epsilon\rightarrow 0}\left[
  \int\limits_0^{\mu-\epsilon}\,+
  \int\limits_{\mu+\epsilon}^\infty\right]
  dE\,\rho(E,\vec{Q})\,n(E,\vec{Q},\vec{x})
\;\end{equation}
and then taking the $\vec{x}$-derivative.
The same decomposition into $T$ and $\mu$-gradient terms
then holds for the currents, i.e.,
with transport coefficients $L_{ij}$ we obtain \cite{KKER86}
\begin{eqnarray}\label{coef}
\vec{j}_l(t,\vec{x})& =
- L_{11} \nabla\mu &- L_{12} \frac{\nabla T}{T}
  \nonumber \\
\vec{E}_l(t,\vec{x})& =
- L_{21} \nabla\mu &- L_{22} \frac{\nabla T}{T}
\;.\end{eqnarray}
Note, that in our picture $L_{12}$ and $L_{21}$ can be different,
i.e. the Onsager relation $L_{ij}=L_{ji}$ is not necessarily fulfilled.
The reason for this is clear: in contrast to ordinary many-body quantum
physics, the present formulation exhibits dissipation already on the
tree-graph level \cite{L88}. In other words,
the formulation of an interacting
theory with continuous spectral functions is {\em not}
micro-reversible,
physical states have a finite lifetime.

In summary of the above reasoning, we can thus write the kinetic
coefficients with the abbreviations
\begin{eqnarray}\label{SDEF}
X_{Ql} &=&   \int\!\!dE\,\left(\rho_l(E,\vec{Q})\right)^2\nonumber \\
Y_{Ql} &=&   \int\!\!dE\,E\,\left(\rho_l(E,\vec{Q})\right)^2
\;,\end{eqnarray}
plus the definitions of
$I_{Ql}$ from eqn. (\ref{npr}), $Z_{Ql}$ from eqn. (\ref{norm})
and $\vec{Q}^{(i)}\vec{Q}^{(j)}\rightarrow Q^2/3$ in very compact form as
\begin{eqnarray}\label{lll}
L_{11} &=&  -2\pi\,\int\!\!\frac{d^3\vec{Q}}{(2\pi)^3}\,
   \frac{ Q }{6} \,\sum\limits_{l=\pm}\,\mbox{sign}(l)\,
        \frac{X_{Ql}}{Z_{Ql}^2}\,
                       \frac{\partial}{\partial \mu}\,
                       \frac{\partial}{\partial Q}\,I_{Ql}
                 \nonumber \\
L_{12} &=&  -2\pi\,\int\!\!\frac{d^3\vec{Q}}{(2\pi)^3}\,
   \frac{ Q }{6}  \,\sum\limits_{l=\pm}\,\mbox{sign}(l)\,
        \frac{X_{Ql}}{Z_{Ql}^2}\,
                       T\frac{\partial}{\partial T}\,
                       \frac{\partial}{\partial Q}\,I_{Ql}
                 \nonumber \\
L_{21} &=&  -2\pi\,\int\!\!\frac{d^3\vec{Q}}{(2\pi)^3}\,
   \frac{ Q }{6}  \,\sum\limits_{l=\pm}\,
                \frac{Y_{Ql}}{Z_{Ql}^2}\,
                       \frac{\partial}{\partial \mu}\,
                       \frac{\partial}{\partial Q}\,I_{Ql}
                 \nonumber \\
L_{22} &=&  -2\pi\,\int\!\!\frac{d^3\vec{Q}}{(2\pi)^3}\,
   \frac{ Q }{6} \,\sum\limits_{l=\pm}\,
                \frac{Y_{Ql}}{Z_{Ql}^2}\,
                       T\frac{\partial}{\partial T}\,
                       \frac{\partial}{\partial Q}\,I_{Ql}
\;.\end{eqnarray}
By subtracting the convective part from the energy current, the
thermal conductivity for the interacting scalar field is obtained
as
\begin{equation}
\lambda = \frac{1}{T}\,\left( L_{22} -\frac{L_{12}L_{21}}{L_{11}}\right)
\;.\end{equation}
$L_{11}=d$ is the diffusion coefficient, and in analogy to
electromagnetic plasmas we call the quantity
\begin{equation}
\alpha = \frac{1}{T}\,\frac{L_{21}}{L_{11}}
\end{equation}
the thermo-force on the charged bosons.

The above formal results can be compared to different calculations
on a numerical as well as a formal level. To begin with the latter,
we insert a simple spectral function for the boson field
at zero chemical potential,
\begin{equation}\label{simple}
\rho_+(E,\vec{k}) = \rho_-(E,\vec{k})  = \frac{2E\gamma_k}{\pi}\,
   \frac{1}{(E^2-\Omega^2_k)^2+4 E^2 \gamma^2_k}
\;\end{equation}
with $\Omega_k^2=\epsilon_k^2+\gamma_k^2$ and
$\gamma_k\ll\epsilon_k$. Such a spectral function can be considered
the lowest order approximation to a system with nontrivial
self energy function \cite{ONSH}.

In ref. \cite{h93pion}, this prescription was applied to the pion
dispersion relation in nuclear matter by rewriting the inverse retarded
propagator for the pions with real energy $E$ as
\begin{equation}\label{inr}
E^2 - E^2_\pi(\vec{k}) - \Pi^R(E,\vec{k})
  = \left( E -( \varepsilon_k-\mathrm{i}\gamma_k)\right)
    \left( E +( \varepsilon_k+\mathrm{i}\gamma_k)\right)
\;.\end{equation}
With this on-shell approximation the integrals are
\begin{eqnarray}
Z_k =
\int\!\!dE\,\rho_l(E,\vec{k})
&=&\frac{1}{2\pi\epsilon_k}\left(\frac{\pi}{2}-\arctan
  \left(\frac{\gamma_k^2-\epsilon_k^2}{
              2\gamma_k\epsilon_k}\right)\right)
  \nonumber\\
&\approx&\;\frac{1}{2\epsilon_k}-\frac{\gamma_k}{\pi\epsilon_k^2}
 +\frac{\gamma^3_k}{3\pi\epsilon^4_k} + {\cal O}(\gamma^4)
\;,\end{eqnarray}
and
\begin{eqnarray}
\frac{X_{kl}}{Z_{kl}^2}&=&
  \frac{1}{2\pi\,\gamma_k} +
  \frac{2}{\pi^2\,\epsilon_k}+
{\cal O}(\gamma) \nonumber \\
\frac{Y_{kl}}{Z_k^2}&=&
  \frac{\epsilon_k}{2\pi\,\gamma_k} +
  \frac{2}{\pi^2}+
{\cal O}(\gamma)
\;.\end{eqnarray}
In the absence of a chemical potential, $L_{12}$ and
$L_{21}$ are zero, but diffusion coefficient and
thermal conductivity are
to lowest order in the width
\begin{eqnarray}\label{lam1}
d& =& -\frac{1}{T}  \,\int\!\!\frac{d^3\vec{k}}{(2\pi)^3}\,
   \frac{ 1 }{3 \gamma_k}\,\frac{\partial}{\partial k}\,
        \epsilon_k\,n(\epsilon_k)\left(1+n(\epsilon_k)\right)
\nonumber \\
\lambda& =& -\frac{1}{T}  \,\int\!\!\frac{d^3\vec{k}}{(2\pi)^3}\,
   \frac{ k\,\epsilon_k }{3 \gamma_k}\,\frac{\partial}{\partial k}\,
        \epsilon_k\,n(\epsilon_k)\left(1+n(\epsilon_k)\right)
\;.\end{eqnarray}
Here, $n(\epsilon_k)$ is the local Bose function taken at energy
$\epsilon_k$.
Apart from the momentum factors, a similar representation
for $\lambda$ was obtained in ref. \cite{HST84}. The
difference can be attributed to the fact that
in this reference a hydrodynamical
picture was assumed together with $\beta\gamma_k\ll 1$.
We thus make a comparision in the high temperature limit, where
one obtains
\begin{equation}\label{asy}
   \lim_{T\rightarrow\infty} \lambda = \frac{A}{\bar{\gamma}}\,T^3
\;.\end{equation}
Here, $\bar{\gamma}$ is some momentum averaged width and $A$
a numerical factor. The same limiting behaviour is obtained in
\cite{HST84}.  Also other calculations
in quantum field theory give qualitatively similar
results: The bulk viscosity of matter
should rise with the same temperature dependence as the thermal
conductivity, and in \cite{T91} (and references quoted there)
was found to rise as $T^3$ in a quark-gluon plasma.

The above expressions are also in accordance
with naive expectations: The thermal conductivity and
diffusion coefficient diverge, if
the width of the particles is reduced, i.e., if the interaction
is removed. In this case the time needed for the relaxation
of a temperature disturbance is infinite.
\section{Application to pion propagation in nuclear matter}
In the next step one has to look at numerical results.
In the spirit of the introductory remarks, pions were chosen for
this example. However, to obtain meaningful results one has to
go beyond the simple approximation discussed above. Instead
of the on-shell approximation of eqn. (\ref{inr}),
the so-called $\Delta$-hole model at finite temperature
is used \cite{h93pion}.

In this model, pions couple to excitations in a gas of (non-relativistic)
nucleons and $\Delta_{33}$ resonances of width $\Gamma$.
Although a numerical calculation of the pionic spectral function
in this system is not very difficult, for the present calculation we
use an analytical approximation
\begin{equation}\label{asys}
\rho_+(E,\vec{k}) = \rho_-(E,\vec{k}) =
\frac{\vec{k}^2C}{\pi}\,\frac{\Gamma\,E\,\omega_\Delta}{
   (E^2-\omega^2_+)^2\,
   (E^2-\omega^2_-)^2 +  \Gamma^2\,E^2\,(E^2-E^2_\pi)^2}
\;.\end{equation}
$E_\pi^2 = \sqrt{\vec{k}^2+m_\pi^2} $ is the free on-shell
energy of the pion,
$\omega_\Delta$ and $\omega_\pm$ are momentum dependent,
\begin{eqnarray}\label{odf}
\omega_\Delta&  =& E_\Delta(\vec{k}) -M_N
    = \sqrt{\vec{k}^2+M_\Delta^2}-M_N \nonumber \\
E_{N\Delta} & = & \vphantom{\int}\sqrt{
     \omega_\Delta
     \left( \omega_\Delta + g^\prime\,C\right)}\nonumber \\
\omega^2_\pm & = & \frac{1}{2}\left(E_{N\Delta}^2
 +(\Gamma/2)^2+E_\pi^2 \pm
  \sqrt{ \left(E_{N\Delta}^2+(\Gamma/2)^2
-E_\pi^2\right)^2 + 4\vec{k}^2 C \omega_\Delta}
  \right)
\;,\end{eqnarray}
and $C$ is the temperature dependent effective coupling constant
(see table)
\begin{table}[b]
\hrule
\vspace*{2mm}
\begin{center}
\begin{tabular}{cccccc} \hline
{}~$f^\pi_{N\Delta}$~ & ~$g^\prime$~ & ~$m_\pi$~ & ~$M_N$~     &
         ~$M_\Delta$~ & ~$\Gamma$~ \\ \hline
         2          & ~0.5~        & ~0.14 GeV~& ~0.938 GeV~ &
          ~1.232 GeV~ & ~0.12 GeV~  \\ \hline
\end{tabular}
\end{center}
\caption{Coupling constants and masses used in the calculations of
 this work.}
\end{table}
  \begin{equation}
C  =
  \frac{8}{9}\left(\frac{f^\pi_{N\Delta}}{m_\pi}\right)^2\,
  \left( \rho^0_N-\frac{1}{4}\rho^0_\Delta\right)\nonumber\\
\;\end{equation}
with partial densities $\rho_N^0$ and $\rho_\Delta^0$ of nucleons
and $\Delta$'s. The partial densities at a given total
baryon density $\rho_B=\rho_N^0 + \rho_\Delta$ are obtained by
a self-consistent calculation of the chemical potential
$\mu_N=\mu_\Delta$ of the fermionic sector.

The above spectral function is obtained with two approximations:
First, its is correct only to lowest order in the baryon
density (quasistatic approximation), and second it is correct
only to lowest order in $\Gamma/M_\Delta$ (asymptotic expansion
of the $\Delta$-spectral function for constant $\Gamma$). These assumptions
yield a polarization tensor of the $\Delta$-nucleon gas
\begin{equation}
\label{ga1}
\Pi^R(E,\vec{k}) = \vec{k}^2C\,\frac{\omega_\Delta}{
                         (E+\mathrm{i}\Gamma/2)^2-\omega_\Delta^2}
\;, \end{equation}
which is then subject to a Migdal (or $g^\prime$) correction
(see refs. \cite{h93pion,EWC93,KXS89} for more details on
this model). Figure \ref{specf} is a logarithmic contour
 plot of the above
spectral function at a total baryon density of $\rho_b$=1.69 nuclear
matter density and temperature $T=0.1 GeV$. It exhibits two branches,
corresponding to the mixing of a coherent $\Delta$-hole excitation
and the pion. At low momenta, the
strength of the excitations lies predominantly on the
lower branch, while at higher momenta the pionic strength has
changed to the upper branch.

In figure \ref{lambdaf}, the product of thermal conductivity $\lambda$
and temperature $T$ which
one obtains with this spectral function is plotted as function
of temperature at different densities. Figure \ref{diffcf}
is a similar representation of the diffusion coefficient $d$.

A comparison of these values to {\em experimental\/}  data is beyond
reach for the time being, we thus have to restrict the
comparison to other calculations. However,
only very few calculations based on field theoretical methods
are available. We have already mentioned the agreement with
ref. \cite{HST84} on a formal level, if a simple approximation
for the spectral function is inserted (eqn. (\ref{simple})).
Although the structure of the more sophisticated spectral
function from (\ref{asys})
is quite different, the high temperature behaviour at the example
density of $\rho_b =1.69\times$ nuclear matter density
inferred from figure \ref{lambdaf} is
\begin{equation}
\lambda \approx 9897\,\mbox{GeV/fm$^2$}\,
\times\,\left[T/\mbox{GeV}\right]^{3.70}
\;,\end{equation}
i.e., it rises even faster than estimated by eqn. (\ref{asy}).
This is due to the decreasing width of the pion with temperature
in the $\Delta$-hole model \cite{h93pion} -- a fact, which was
not included in eqn. (\ref{asy}). At lower temperatures,
the thermal conductivity rises slower with temperature than the
asymptotic expression.

The density dependence of the coefficients $\lambda$ and $d$
is quite small, with the general tendency to have a ''stiffer''
temperature dependence at lower baryon density.
This is again consistent with the fact, that the pion in the $\Delta$-hole
model becomes a free particle without the baryonic background.

The exploratory calculation of the present paper was done
with a constant width $\Gamma$ for the $\Delta$-resonance. Even
the vacuum spectral function of the $\Delta$ is not completely
described by such an ansatz, much less its medium
dependence \cite{KXS89,K86}. The width in matter at finite
temperature (obtainable by self-consistent calculations currently in
progress) might be a factor 2--3 times higher, hence according
to eqn. (\ref{lam1}) the thermal conductivity should be
lower due to the medium dependence of the $\Delta$-width,
and its temperature dependence flatter.

For completeness, also {\em variational\/} and
{\em hydrodynamical \/} calculations of the pionic thermal
conductivity
in the baryon-free regime have to be mentioned \cite{G85,W93}
(also plotted in figure \ref{lambdaf}). In these,
the thermal conductivity for $T\stackrel{>}{\sim}$ 0.1 GeV
is much lower than
obtained in the present paper: Beyond $T\approx$ 60 MeV
it rises only quadratically with
temperature (see thin lines in figure \ref{lambdaf} for
the corresponding slopes). The results have been obtained
by using vacuum pion-pion scattering data as input, and thus
the interaction in these models
is completely different from the $\Delta$-hole model.
Hence, a direct comparision of the results is not
justified.

Nevertheless, a short remark on the difference is necessary.
Within the hydrodynamic picture a medium effect, and thus also
the $T^2$-dependence of the thermal conductivity, can be explained
by the rise of the effective pion-pion scattering cross-section
with temperature \cite{BGL88}. It was already stated,
that the medium broadening of the pion might be
underestimated in the present paper. However,
even when reducing the present results by a factor 2--3 at
higher temperatures, we still find that the
thermal conductivity of the interacting pions in presence of
the hot baryonic background is higher than in refs. \cite{G85,W93}.
\section{Summary and Conclusions}
In the present paper, Thermo Field Dynamics (TFD) for inhomogeneous
systems \cite{Ubook,NUY92} is generalized to quantum fields
with a continuous single-particle mass spectrum \cite{L88}.
Following a procedure outlined in refs. \cite{YUNA92,hu92,hu92a},
the spectral and statistical content of the full
two-point function are separated in a well defined approximation.
This approximation, i.e., assuming that a {\em spectral function\/} still
is a meaningful concept, limits the results to situations not too
far from equilibrium. However, corrections to this approximation
are of second order in the gradients. Furthermore a general formulation
in coordinate space is easily achieved \cite{YUNA92,hu92},
which also takes temporal inhomogeneity into account and thus does
not trade simplicity for covariance.

By requiring the thermal Bogoliubov symmetry to hold locally in space,
a modification of the systems' hamiltonian was obtained.
It is well known, that this {\em gauging\/} of
a global symmetry introduces a minimal coupling to an external
classical field, here: A ''temperature field'' and a
''chemical potential field'' \cite{h90ber}.

In a system close to equilibrium, this coupling was used to calculate
transport (or kinetic) coefficients of the interacting quantum field.
In contrast to previous derivations \cite{HST84}, neither imaginary time
arguments nor perturbation theory in powers of a coupling constant
were used for this purpose.
Expressions for thermal conductivity and the diffusion coefficient
were given in terms of the equilibrium spectral function of the
quantum field. They are dominated by two factors: A Poisson bracket,
which stems from a gradient expansion (cf. eqn. (\ref{cur})), and
proportionality to the inverse width of the spectral distribution,
cf. eqn. (\ref{lam1}).

The Poisson bracket contributes a factor $\hbar$, while in a naive
counting the inverse width contributes a factor $1/\hbar$ to the
transport coefficient. This cancellation
seems to be the main source for the difficulties one has
in obtaining kinetic coefficients from quantum field theory
\cite{E86,D84a}.

A different approach to the calculation of transport coefficients
is the solution of a Boltzmann equation (e.g. in relaxation time
approximation) \cite{G85,M93}. However, this approach has
the drawback of a zero width quasi-particle approximation at a very
early stage of the computation, i.e., it is semi-classical.
In contrast to such kinetic models,
the formulation of the present work contains {\em off-shell effects\/},
and therefore the full quantum description, in a consistent way.

Furthermore, some of the semi-classical calculations
retain the relaxation time as an undetermined
factor in the final expressions. While TFD in principle also
provides an expression for the relaxation time \cite{YUNA92,hu92},
the method presented here is more direct and can be
extended systematically.

The new method was tested by calculating thermal conductivity and
diffusion coefficient for the pionic component in a hot $\Delta$-nucleon
gas. Results were obtained, which are substantially
higher than those found with the semi-classical zero width
kinetic picture. As was pointed out in the previous section,
part of this might be due to the too
simplistic picture of the $\Delta$-hole model with constant
$\Delta$-width.

However, this finding is also consistent
with an observation made, if the Kadanoff-Baym
equations are solved directly rather than approximated
by a Boltzmann equation: The relaxation of perturbations
is slower, if the full quantum description is taken into account
\cite{D84}.

\subsection*{Acknowledgements}
I express my thanks to H.Umezawa, B.Friman and R.Sollacher for
useful comments.
\clearpage

\begin{figure}
\vspace*{165.4mm}
\includegraphics{spec.ps}
\caption{Pion spectral function of the $\Delta$-hole model
         at $\rho_b$ = 1.69 nuclear density and $T=0.1$ GeV.}
\label{specf}
\small
Horizontal axis $|\vec{k}|$ in GeV, vertical axis $E$ in GeV.
Plotted are lines of equal $\rho(E,\vec{k})$ on\\ logarithmic scale,
see ref. \cite{h93pion} for details.
\normalsize
\end{figure}
\begin{figure}
\vspace*{165.4mm}
\includegraphics{lambda.ps}
\caption{Pion thermal conductivity $\times$ temperature.}
\label{lambdaf}
\small
Thin straight lines: slopes $\propto T^3$, $T^4$\\[1mm]
Full thick line: $\Delta$-hole model at $\rho_b$ = 1.69 nuclear density;\\
dashed line 0.92 and dash-dotted line 0.47 nuclear density.\\[1mm]
Dash-double-dotted line ref. \cite{W93}, dotted line
ref. \cite{G85},\\
 both from $\pi-\pi$ scattering data at $T=0$, $\rho_b=0$.
\normalsize
\end{figure}
\begin{figure}
\vspace*{165.4mm}
\includegraphics{diff.ps}
\label{diffcf}
\small
\caption{Pion diffusion coefficient.}
Thin straight lines: slopes $\propto T^3$, $T^4$\\[1mm]
Full thick line: $\Delta$-hole model at $\rho_b$ = 1.69 nuclear density;\\
dashed line 0.92 and dash-dotted line 0.47 nuclear density.\\[1mm]
\normalsize
\end{figure}

\begin{thebibliography}{99}
\bibitem{L88}{
    N.P.Landsman, Ann.Phys. {\bf 186 } (1988) 141}
\bibitem{Ubook}{
    H.Umezawa,
    {\em Advanced Field Theory: Micro, Macro and Thermal Physics},\\
    (American Institute of Physics, 1993)}
\bibitem{DP91}{
   J.E.Davis and R.J.Perry, Phys.Rev.{\bf C 43} (1991) 1893}
\bibitem{BS75}{
    H.J.Borchers and R.N.Sen,
    Commun.Math.Phys.{\bf 21} (1975) 101}
\bibitem{YUNA92}{
    H.Umezawa and Y.Yamanaka, Mod.Phys.Lett {\bf A7} (1992)  3509;\\
    Y.Yamanaka, H.Umezawa, K.Nakamura and T. Arimitsu,\\
    {\em Thermo Field Dynamics in Time Representation},\\
    University of Alberta Preprint (1992)}
\bibitem{hu92}{
    P.A.Henning and H.Umezawa,\\
    {\em Diagonalization of Propagators in TFD
     for Relativistic Quantum Fields}\\
    GSI-Preprint 92-61 (1992), subm. to Nucl.Phys. {\bf B}}
\bibitem{hu92a}{
    P.A.Henning and H.Umezawa, Phys.Lett. {\bf B 303} (1993) 209}
\bibitem{HST84}{
    A.Hosoya, M.Sakagami and M.Takao,
    Ann.Phys. {\bf 154} (1984) 229}
\bibitem{KKER86}{
    W.D.Kraeft, D.Kremp, W.Ebeling and G.R\"opke,\\
    {\em Quantum Statistics of Charged Particle Systems\/},\\
    (Plenum Press, New York 1986)}
\bibitem{K57}{
    R.Kubo, J.Phys.Soc. Japan {\bf 12} (1957) 570}
\bibitem{TBAB91}{
    K.Tanaka, W.Bentz, A.Arima and F.Beck,
    Nucl.Phys. {\bf A528} (1991) 676}
\bibitem{h92fock}{
    P.A.Henning, Nucl.Phys. {\bf A546} (1992) 653}
\bibitem{X93}{
    Hong-Hua Xu, Phys.Rev {\bf D 47} (1993) 2622}
\bibitem{NUY92}{
    K.Nakamura, H.Umezawa and Y.Yamanaka,
    Mod.Phys.Lett {\bf A7} (1992)  3583}
\bibitem{h90ber}{
    P.A.Henning, M.Graf and F. Matth\"aus,
    Physica {\bf A 182} (1992) 489}
\bibitem{H90}{
    P.A.Henning, Nucl.Phys. {\bf B337} (1990) 547,
                 Phys.Lett. {\bf A145} (1990) 329}
\bibitem{ONSH}{
    H.Chu and H.Umezawa\\
   {\em Stable quasi-particle picture in thermal quantum field physics\/},\\
    University of Alberta Preprint (1992)}
\bibitem{h93pion}{
    P.A.Henning and H.Umezawa,\\
    {\em The Delta-hole model at finite temperature\/},\\
    GSI-Preprint 93-23 (1993), subm. to Nucl.Phys. {\bf A}}
\bibitem{T91}{
    M.Thoma, Phys.Lett. {\bf B269} (1991) 144}
\bibitem{EWC93}{
    W.Ehehalt, Gy.Wolf, W.Cassing et. al., Phys.Lett. {\bf B298} (1993) 31}
\bibitem{KXS89}{
    C.M.Ko, L.H.Xia and P.J.Siemens,  Phys.Lett. {\bf B 231} (1989) 16}
\bibitem{K86}{
    Y.Kitazoe, M.Sano, H.Toki and S. Nagamiya,
    Phys.Lett. {\bf B 166} (1986) 35}
\bibitem{G85}{
    S.Gavin, Nucl.Phys. {\bf A435} (1985) 826}
\bibitem{W93}{
    M.Prakash, R.Venugopalan and G.Welke,
    {\em Non-equilibrium properties of hadronic mixtures},\\
    Wayne State University preprint (1993), to appear in Phys.Rep.}
\bibitem{BGL88}{
    G.Bertsch et.al., Phys.Rev. {\bf D 37} (1988) 1202}
\bibitem{E86}{
    H.Ezawa, {\em in\/}
    Progress in Quantum Field Theory, \\
    Eds. H.Ezawa and S.Kamefuchi (Elsevier 1986)}
\bibitem{D84a}{
    P.Danielewicz,
    Phys.Lett. {\bf B146} (1984) 168}
\bibitem{M93}{
    L.Mornas,\\
    {\em Collective Effects on Transport Coefficients of Relativistic
    Nuclear Matter\/},\\
    GSI-Preprint 93-21 (1993)}
\bibitem{D84}{
    P.Danielewicz,
    Ann.Phys. {\bf 152} (1984)  239 and 305}
\end{thebibliography}
\end{document}